\def\cps{s$^{-1}$}

\documentstyle[aaspp4]{article}

\begin{document}

\title{Broad-band X-Ray Spectra of the Black Hole Candidate GRO~J1655--40}

\authoremail{jat@astro.columbia.edu}

\author{John A. Tomsick\altaffilmark{1}, Philip Kaaret\altaffilmark{1,4}, Richard A. Kroeger\altaffilmark{2}, Ronald A. Remillard\altaffilmark{3}}
\altaffiltext{1}{Department of Physics and Columbia Astrophysics Laboratory, Columbia University, 550 W. 120th Street, New York, NY 10027}
\altaffiltext{2}{Naval Research Laboratory, Code 7650, Washington DC}
\altaffiltext{3}{Center for Space Research, Massachusetts Institute of Technology, Cambridge, MA 02139}
\altaffiltext{4}{Current address: Harvard-Smithsonian Center for Astrophysics, 60 Garden Street, Cambridge, MA 02139 (e-mail: pkaaret@cfa.harvard.edu)}

\begin{abstract}

We present broad-band (2~keV to 2~MeV) x-ray spectra of GRO~J1655--40,
a luminous X-ray transient and occasional source of relativistic
radio jets, obtained with the \it Rossi X-Ray Timing Explorer \rm 
(RXTE) and the Oriented Scintillation Spectrometer Experiment (OSSE).  
In one observation, the luminosity is found to be 18\% of the Eddington
limit, which is one of the highest luminosities ever observed from
GRO~J1655--40.  For this observation, we find that an adequate fit is
obtained when a broad iron line and a reflection component are added
to a model consisting of a power-law plus a soft excess component.
The 95\% confidence lower limit on the rms line width is 0.86~keV.
The power-law component has a photon index of $2.72^{+0.03}_{-0.08}$ 
and extends to at least 800~keV without a cutoff.

After this observation, a significant drop in the (5-12~keV)/(1.5-5~keV)
hardness ratio occured on a timescale less than 2 hours.  From an RXTE
observation of GRO~J1655--40 made after the hardness transition, we
find that the power-law index is harder ($\alpha = 2.415\pm 0.011$), 
the flux of the power-law component is lower, and the total luminosity
is 10\% of the Eddington limit.  The change in the power-law component 
is consistent with the correlation between the spectral index and power-law 
flux previously reported for GRO~J1655--40.

\end{abstract}


\clearpage

\section{Introduction}

The X-ray transient GRO~J1655--40 was discovered with the Burst and Transient
Source Experiment (BATSE) on the \it Compton Gamma-Ray Observatory \rm
(CGRO) on July 27, 1994 (\cite{zhang94}).  GRO~J1655--40 is often called a 
microquasar because radio jets moving at 92\% of the speed of light relative to 
the X-ray source have been observed (\cite{tingay95};~\cite{hjellming95}).  From 
optical measurements made in X-ray quiescence, the compact object mass and the orbital 
inclination have been found to be $7.02\pm 0.22 M_{\sun}$ and $69.50\pm 0.08$~degrees, 
respectively (\cite{orosz97}).  These are the most precise compact object 
mass and orbital inclination measurements which have been made for any black hole
candidate, which makes GRO~J1655--40 an excellent laboratory for learning about 
accreting black holes.

Since the discovery of GRO~J1655--40, its 20-200~keV X-ray flux has been monitored 
by BATSE (\cite{tavani96}).  In the BATSE bandpass, a power-law spectrum
is observed with a photon index varying between 1.8 and 3.1.  The BATSE data
suggest that the photon index softens as the source intensity increases 
(\cite{wilson95};~\cite{zhang95}).  This trend has also been reported from 
observations of GRO~J1655--40 by the Oriented Scintillation Spectrometer Experiment 
(OSSE) on CGRO (\cite{grove98}).  OSSE measurements of GRO~J1655--40 have been made 
between August 1994 and September 1996.  During observations where the source was 
detected, the spectrum in the OSSE bandpass is consistent with a power-law 
(\cite{kroeger96};~\cite{grove98}).  In this paper, we report on simultaneous RXTE 
and OSSE observations of GRO~J1655--40 made in 1996.  Soft X-ray measurements of 
the source made with the \it Advanced Satellite for Cosmology and Astrophysics \rm 
(ASCA) indicate the presence of an ultrasoft component (\cite{zhang97a}),
which probably comes from an accretion disk around the compact object.

Previous spectral and timing measurements of GRO~J1655--40 have been interpreted
as evidence that the compact object in this system is a rapidly rotating 
black hole (Zhang, Cui \& Chen 1997b;~\cite{cui98a}).  Similar results have been 
obtained for another microquasar, GRS~1915+105, suggesting that the relativistic, 
highly collimated radio jets in the microquasars are related to the presence of a 
rapidly rotating black hole (\cite{zhang97b}).  Black hole rotation rate has long 
been considered as a way to explain the radio loud/radio quiet dichotomy 
in AGN (\cite{wc95}).

In this paper, we report on simultaneous observations of GRO~J1655--40
made by the \it Rossi X-Ray Timing Explorer \rm (RXTE) and OSSE.  After 
describing the observations and our analysis techniques, we present
the results of the spectral fits.  We then discuss the results and end
with our conclusions.

\section{Observations}

Figure~1 shows the GRO~J1655--40 RXTE All-Sky Monitor (ASM) light curve and
(5-12~keV)/(1.5-5~keV) hardness ratio from late August to early September 1996 when
GRO~J1655--40 was observed simultaneously with RXTE and OSSE.  OSSE observations were 
made following observations by BATSE indicating that GRO~J1655--40 had undergone several 
strong outbursts since mid-July 1996.  The decision to observe GRO~J1655--40 came during 
the fifth outburst of this series above 1 Crab in the 20-100~keV energy band.  These 
RXTE and OSSE observations occur during one of the highest luminosity and hardest 
states of GRO~J1655--40 (\cite{remillard98}).  A sharp change in the ASM hardness is seen 
at MJD 50328.5, which is marked with a vertical dotted line in Figure~1.  
For the six days before the transition, the ASM count rate is constant (in 1-day averages)
to $\pm6$\% with a mean value of $220\pm 2$~\cps, corresponding to about 2.9~Crab. 
Over the same time period, the mean hardness ratio is $0.449\pm 0.003$.  
At the transition, the hardness changes on a timescale less than 2~hours, while the 
1.5-12~keV flux dips sharply, recovers, and then decays to a constant level in
about 1.5~days.  During the time from MJD 50330 to MJD 50335, the mean ASM count 
rate is $170\pm 1$~\cps~and the one day averages vary by $\pm4$\%.  The OSSE observation 
was made from MJD 50322.4 to MJD 50328.4 and lies completely in the time interval of high 
X-ray hardness.  Pointed RXTE observations, marked with arrows in Figure~1b, were made on 
MJD 50324.4 and MJD 50330.3.  One RXTE observation is during the OSSE observation and the 
other is after the hardness transition.  The first RXTE observation contains 4334~s of PCA 
livetime and 2929~s of HEXTE livetime (on source).  The second RXTE observation contains 
5618~s of PCA livetime and 3715~s of HEXTE livetime.

GRO~J1655--40 radio observations were made with the \it Very Large Array \rm (VLA) and the 
\it Molonglo Observatory Synthesis Telescope \rm (MOST) close to the time of the RXTE and 
OSSE observations (R.M. Hjellming \& R. Hunstead, private communication).  GRO~J1655--40 
was not detected for any of the radio observations.  Observations were made by the VLA on 
MJD 50323, 50326, 50330, and 50337 resulting in upper limits of 0.4~mJy, and observations 
were made by MOST on MJD 50306, 50313, 50326, and 50340 resulting in an upper limit of 4~mJy.

\section{Analysis}

PCA energy spectra have been produced using Standard~2 PCA data and the Version 2.2.1
(January 20, 1998) response matrix (\cite{jahoda97}).  During the pre-transition observation, 
Proportional Counter Unit (PCU) 3 was off.  To test the PCA response matrix for the remaining 
PCUs (0, 1, 2, and 4), we fit the Crab spectrum with a power-law.  Crab spectra 
have been extracted from an 11.6~ks observation made between MJD 50529.9 and 50530.2.  
We produced spectra containing data from all three xenon layers and also from
only the top xenon layer.  In both cases, power-law fits to the Crab are significantly
better for PCUs 0, 1, and 4 than for PCU~2.  Also, the top layer spectra give column 
densities and photon indices which are closer to the conventional values for the
Crab (\cite{toor74};~\cite{koyama84};~\cite{schattenburg86};~\cite{turner89}) than the 
spectra for all three layers combined.  Thus, the GRO~J1655--40 spectral analysis has
been performed using events from the top layers of PCUs 0, 1, and 4.
Although PCU~3 was on during the post-transition observation, it has not been used in
order to avoid instrumental differences between the 
pre-transition and post-transition spectra.  A systematic error of 1\% is assumed to 
account for uncertainties in the response matrix, and the normalizations have been left 
free between PCUs.  Spectral bins from 2.5~keV to 20.0~keV have been used for the
spectral analysis.  Background subtraction has been performed including estimates
for the particle, X-ray, and activation background spectra (\cite{stark97});  however, 
for both spectra, activation does not contribute to the background.  The spectra
have been corrected for deadtime.  

We find that pile-up has a significant impact on the GRO~J1655--40 PCA spectrum.
The ratio of the pile-up spectrum to the measured spectrum reaches a maximum of 5\% 
near 18~keV for the pre-transition spectrum, while the ratio reaches a maximim of
5\% near 13~keV for the post-transition spectrum.  Thus, a correction for pile-up 
has been applied as described in Tomsick \& Kaaret (1998).

HEXTE energy spectra have been produced using Standard mode data, which consist of
64-bin spectra with 16~s time resolution.  The March~20,~1997 HEXTE response matrices
have been used.  For HEXTE, it is essential to correct for deadtime
(\cite{rothschild97}), and this correction has been made.  For the spectral fits, 
the normalizations have been left free between cluster A and cluster B.  

OSSE made its standard observation over a period of six days by pointing its collimator 
at GRO~J1655--40, and comparing this with count rates in background fields 4.5 degrees 
on either side of the source.  Due to the large field of view of OSSE, 
$3.8^{\circ}$ by $11.4^{\circ}$ full width at half maximum response, the observations 
have potential problems with source confusion.  Nearby sources 4U~1700--37 and 
OAO~1657--415 are not significant sources above 100~keV in comparision to GRO~J1655--40 
and can be ignored.  To make sure that the OSSE data is not contaminated due to source 
confusion, we compared the OSSE spectrum to the HEXTE spectrum in the energy band where 
the two instruments overlap.  We find that the OSSE and HEXTE spectra are consistent, 
up to a constant normalization factor, in this energy band.  Narrow-line annihilation 
radiation from the galactic plane could potentially be a source of contamination; 
however, for this observation, the orientation of the spacecraft was chosen to
minimize this contamination.  No evidence for 511~keV line emission from 
GRO~J1655--40 was detected in these observations.  

The normalizations between the PCA, HEXTE, and OSSE have been left free.  
To estimate the GRO~J1655--40 flux, we have compared the Crab nebula flux measured
by the PCA to previous measurements of the Crab flux by other instruments
(\cite{toor74};~\cite{koyama84};~\cite{schattenburg86};~\cite{turner89}).  We find
that PCUs 0, 1, and 4 measure 2.5 to 25~keV Crab fluxes of 3.43, 3.38, and 
3.29 ($\pm 0.05$) photons~cm$^{-2}$~s$^{-1}$, respectively, while previous measurements 
give $2.91\pm 0.03$~photons~cm$^{-2}$~s$^{-1}$ in the same energy band.  When fluxes or 
spectral component normalizations are given in this paper, they have been reduced by a 
factor of 1.18 so that the PCA flux scale is in agreement with previous instruments.
Spectral fits have been calculated using the version 10 XSPEC software (\cite{shafer91}).

\section{Spectral Results}

We first fit the pre-transition GRO~J1655--40 HEXTE and OSSE spectra without 
the data from the PCA.  Since a power-law does not provide an acceptable fit 
($\chi^{2}/\nu$ = 197/140), more complicated models were used to improve the fit.  The 
fit results are shown in Table~1.  A model consisting of a power-law with an exponential 
break (cutoff power-law) gives $\chi^{2}/\nu$ = 149/139, which is a significant 
improvement over a power-law.  The photon index and folding energy are found to be 
$2.703\pm 0.010$ and $1086^{+195}_{-145}$~keV, respectively.  The error estimates 
given here and throughout the text are 68\% confidence for one interesting 
parameter ($\Delta \chi^{2} = 1.0$).  A similar improvement is observed using a model 
consisting of a power-law with a reflection component (\cite{magdziarz95}).  For this 
fit, we fixed the inclination angle to 69.5\arcdeg, which is the binary inclination 
angle measured by Orosz \& Bailyn (1997), and assumed that the reflecting material has 
cosmic abundances.  We find that the photon index and covering fraction 
($\Omega/2\pi$) are $2.739\pm 0.006$ and $0.396\pm 0.065$, respectively, and that the 
spectral shape is consistent with the reflecting material being unionized.  A broken 
power-law with a break at $63^{+9}_{-14}$~keV also provides a good fit to the data.  
The photon indices for this fit are shown in Table~1.

When the PCA data is added to the pre-transition spectrum, a soft excess is observed, 
which is probably due to emission from an accretion disk.  We first use a disk-blackbody 
component (\cite{makishima86}) to model the soft excess.  The disk-blackbody model we
use is not standard in XSPEC version 10.  The difference between our model and the 
standard XSPEC disk-blackbody model (``diskbb'') is that we use the relation between 
temperature and radius is given in equation (3.23) of Pringle (1981).  For our model, 
which is described in more detail in Tomsick et al. (1998), the spectrum is determined 
by the normalization, $N_{DBB}$, and the maximum disk color temperature, $T_{max}$.  
The normalization is given by
\begin{equation}
N_{DBB} = \frac{\rm cos\it~i}{f^{4}} \left(\frac{R_{in}/1~\rm km\it}{d/10~\rm kpc\it}\right)^{2},
\end{equation}
where $i$ is the disk inclination, $f$ is the color correction factor, $R_{in}$ is the
inner radius of the disk, and d is the distance to the source.  In the following, we
assume that $f = 1.7\pm 0.2$ (\cite{shimura95};~\cite{zhang97b}) and that 
$d = 3.2\pm 0.2$ (\cite{hjellming95}).  We also model the soft excess using a
Comptonization component (\cite{st80}) rather than the disk-blackbody.  The 
Comptonization model approximates the situation where soft X-rays from the accretion 
disk are upscattered in a plasma above the accretion disk.  It is important to note 
that the Comptonization model we use only applies if the plasma is relatively optically 
thick ($\tau > 3$) and the electrons in the plasma are non-relativistic 
(i.e. $kT_{e} \ll m_{e}c^{2}$).  After the spectra are fit using this Comptonization
model, it will be necessary to check that the fit parameters satisfy these conditions.
A disk-blackbody plus power-law model does not provide a formally acceptable fit to the 
pre-transition spectrum ($\chi^{2}/\nu$ = 391/274).  A Comptonization plus power-law model 
provides a significantly better fit; however, it is still not formally acceptable 
($\chi^{2}/\nu$ = 291/273).  As shown in Figures 2a and 2b, significant systematic features 
are seen in the residuals for both models.

To see if the systematic features in the GRO~J1655--40 residuals are due to problems with 
understanding the instrument responses, we produced PCA and HEXTE Crab spectra.
To make the spectra, we used data from the 11.6~ks Crab observation mentioned previously.  
The Crab spectra were created using the same procedures as for the GRO~J1655--40 data.  
We note that only top layer PCA data are used.  The Crab spectra and the residuals for 
a broken power-law fit are shown in Figure~3.  The data is well fit with a broken 
power-law ($\chi^{2}/\nu$ = 146/221).  The systematic features that are observed in the 
GRO~J1655-40 residuals are not seen in the Crab residuals; thus, we assume that the 
features in the GRO~J1655--40 residuals are not due to problems in understanding the 
instrument responses.

The positive residuals in the pre-transition spectrum between 6 and 8~keV suggest the 
presence of an iron line.  Since an iron line is expected if reflection is being observed, 
we choose to model the high energy portion of the spectrum using a power-law with a 
reflection component.  Thus, in the following, we use a model consisting of a soft 
component (disk-blackbody or Comptonization), a gaussian emission line, and a power-law 
with a reflection component to fit the GRO~J1655--40 spectra.  As shown in Tables 2 and 3, 
similar results are obtained using the disk-blackbody model and the Comptonization model.  
Specifically, we find that both reflection and the gaussian emission line are necessary at 
more than 99\% significance.  Henceforth, Model 1 will refer to a model consisting of a 
disk-blackbody, a gaussian emission line, a power-law, and reflection, and Model 2 will 
refer to a model consisting of a Comptonization component, a gaussian emission line, 
a power-law, and reflection.

The fit parameters for the Model 1 and Model 2 fits to the pre-transition spectrum are
shown in Tables 4 and 5, respectively.  For the reflection component, we measure a 
covering fraction of $0.259^{+0.054}_{-0.053}$ using Model~1 and $0.85^{+0.68}_{-0.25}$ 
using Model 2.  Using Model 1, we find that the iron line centroid ($E_{line}$) is 
$6.81^{+0.24}_{-0.31}$~keV, and that the equivalent width is 113~eV.  For 
Model 2, $E_{line} = 6.21^{+0.31}_{-0.33}$~keV and the equivalent width is 
319~eV.  In both cases, the line is very broad.  The 95\% confidence lower limit 
($\Delta\chi^{2} = 4.0$) on the width of the line ($\sigma_{line}$) is 0.86~keV 
and 1.42~keV for Models 1 and 2, respectively.  The residuals for pre-transition Model 1 
and Model 2 fits are shown in Figures 2c and 2d, respectively, and the pre-transition 
spectrum, fit using Model 1, is shown in Figure~4a.  For both models, the iron line centroid 
does not constrain the ionization state of the reflecting material.  Thus, to constrain the 
ionization state of the material, we freed the ionization paramter ($\xi$) in the 
reflection model and, assuming a disk temperature of $10^{5}$~K (\cite{cui98b}), we 
refit the spectra.  For Model 1 and Model 2, we find that the best value for 
$\xi$ is 0.0, and, for this parameter, we derive 95\% confidence upper limits on $\xi$
of 1380~erg~cm~s$^{-1}$ and 48~erg~cm~s$^{-1}$ for Model 1 and Model 2, respectively.  
Although the spectrum is consistent with the reflecting material being unionized, $\xi$ is 
not very well constrained.  The fact that $\xi$ is not well constrained is not surprising 
because of the presence of the broad iron line.  For Model 2, we find that the conditions 
for the validity of the Comptonization model are met.  Specifically, we find that
$\tau = 7.64^{+0.75}_{-0.65}$ and that the Comptonization component is responsible 
for less than 1\% of the flux for all energies greater than 28~keV, indicating that 
relativistic corrections to this model are not significant in this case.

When the post-transition spectrum is fit with a disk-blackbody plus power-law model, 
the fit is formally acceptable ($\chi^{2}/\nu$ = 187/221).  However, as for the
pre-transition spectrum, we tried to improve the fit using a reflection component
and a gaussian line.  As shown in Table~2, we find that including a reflection component 
improves the fit only slightly.  Adding a gaussian iron line improves the fit significantly, 
and we find that this component is necessary at more than 99\% significance.  Further evidence 
that the reflection component is not necessary to fit the post-transition spectrum comes from 
the fact that adding a reflection component to a model consisting of a disk-blackbody, a 
gaussian iron line, and a power-law does not improve the fit.  As shown in Table~3, similar 
results are obtained using the Comptonization model instead of the disk-blackbody.  Specifically, 
we find that the gaussian iron line is necessary at more than 99\% significance, but there is 
no evidence for reflection.

The fit parameters for the Model 1 and Model 2 fits to the post-transition spectrum are
shown in Tables 4 and 5, respectively.  For the post-transition fits described above, all 
three gaussian iron line fit parameters were left as free parameters.  However, we find that
for both Model 1 and Model 2, the post-transition values found for $E_{line}$ and 
$\sigma_{line}$ are consistent with those found for the pre-transition fits.  Thus, the 
post-transition fits were recalculated after fixing $E_{line}$ and $\sigma_{line}$ to the 
pre-transition values in order to improve the constraints on the other parameters.
For both models, we find that the equivalent width of the iron line is lower for the 
post-transition spectrum than for the pre-transition.  For Model 1, the equivalent width is 
38~eV, and for Model 2, the equivalent width is 192~eV.  We also place upper limits on the 
covering fraction for reflection.  Using Model 1, we find that the 95\% confidence 
upper limit on $\Omega/2\pi$ is 0.067, while using Model 2 the 95\% confidence upper limit is 
0.15.  Thus, reflection is significantly less important in the post-transition spectrum 
compared to the pre-transition spectrum.  Another significant difference between the spectra, 
which will be discussed further below, is that the power-law index for the post-transition 
spectrum is considerably harder than for the pre-transition spectrum.  The post-transition 
spectrum, fit using Model 1, is shown in Figure~4b.

ASCA observations of GRO~J1655-40 give column densities ranging from 
$N_{\rm H} = 4.4\times10^{21}\rm~cm^{-2}$ (Nagase et al.~1994) to 
$N_{\rm H} = 8.9\times10^{21}\rm~cm^{-2}$ (Zhang et al.~1997a).  Fitting the pre-transition 
spectrum with Model 1 gives a column density which is significantly above this range 
(cf. Table 4).  To check this result, we fixed $N_{\rm H}$ to the value found by 
Zhang et al. (1997a) and refit the pre-transition spectrum using Model 1.  With 
$N_{\rm H}$ fixed, we find that $\chi^{2}/\nu$ = 293/272, which is significantly 
worse than the fit with $N_{\rm H}$ free.  The column density for the pre-transition spectrum 
may be higher than the previous values found using ASCA and higher than we find for the 
post-transition spectrum because of absorption due to material near the source.  When Model 2 
is used, the column densities for the pre-transition and post-transition spectra are consistent 
(cf. Table 5).  Although the values of $N_{\rm H}$ found using Model 2 are above the range of 
values found using ASCA, the ASCA column densities are derived using different spectral models.

From ASCA spectra, Ueda et al. (1998) recently reported a detection of iron absorption 
lines for GRO~J1655--40.  In the August 1995 ASCA spectrum, Ueda et al. (1998) find a 
K$\alpha$ absorption line due to highly ionized iron (Fe XXVI) with an equivalent 
width of $25^{+13}_{-11}$~eV.  Ueda et al. (1998) derives a 1-$\sigma$ upper limit on the 
width of the iron line of 150~eV.  To look for this absorption line in our spectra, we 
added a gaussian at a fixed energy of 6.98~keV and a fixed width of 150~eV to Model 1 and
refit the pre-transition and post-transition spectra.  An absorption line is not detected 
for either the pre-transition or post-transition spectrum.  For the pre-transition
spectrum, we find a 95\% confidence upper limit on the equivalent width of 24.4~eV, and for 
the post-transition spectrum, we find a 95\% confidence upper limit on the equivalent width 
of 26.5~eV.  Although we do not detect an iron absorption line, our data does not allow us to 
rule out the absorption line observed by Ueda et al. (1998).  However, it appears that the 
structure between 6 and 8 keV in our spectrum is considerably different from that observed by 
ASCA since we observe an iron emission line that was not detected by ASCA.

\section{Discussion}

For GRO~J1655--40, the change from the pre-transition state to the post-transition state
represents a significant change in the luminosity, and hereafter the pre-transition state
will be referred to as the high-luminosity state and the post-transition state will be referred 
to as the intermediate-luminosity state.  The unabsorbed luminosity for the high-luminosity 
state is $1.7\times 10^{38}$~erg~s$^{-1}$ (1.5 keV to 2 MeV), assuming a distance of 3.2~kpc
and isotropic emission, corresponding to 18\% of the Eddington luminosity ($L_{edd}$) for a 
$7M_{\sun}$ compact object (\cite{orosz97}).  For the intermediate-luminosity state, the 
unabsorbed luminosity is $9.2\times 10^{37}$~erg~s$^{-1}$ (1.5~keV to 2~MeV), which is 
10\% of $L_{edd}$.  Figure~5 shows the high-luminosity and intermediate-luminosity state 
spectra along with a previously reported GRO~J1655--40 broad-band spectrum based on 
simultaneous August 1995 ASCA and BATSE observations (\cite{zhang97a}).  The 
intermediate-luminosity state spectrum is very similar to the August 1995 spectrum, while 
the shape of the high-luminosity state spectrum is much different.

For the power-law component, the intermediate-luminosity state photon index is considerably 
harder than the high-luminosity state photon index ($\alpha = 2.415\pm 0.011$ compared to 
$\alpha = 2.72^{+0.03}_{-0.08}$).  Also, the 2.5~keV to 2~MeV flux of the power-law component 
is about a factor of three higher for the high-luminosity state than the 
intermediate-luminosity state.  The differences in the power-law component between the 
high-luminosity and intermediate-luminosity state are consistent with the trend previously 
reported for GRO~J1655--40 by Wilson et al.~(1995) from BATSE measurements and by 
Grove et al.~(1998) from OSSE measurements that the spectral index softens as the 
intensity of the hard component increases.  

Chakrabarti \& Titarchuk (1995) have suggested that the power-law component may be caused 
by the Comptonization of soft photons from the accretion disk by material being radially
accreted onto the black hole.  This mechanism, known as bulk-motion Comptonization, is 
different from thermal Comptonization models (e.g. \cite{titarchuk94}) since it relies on 
the radial velocity of the infalling material to produce hard x-rays.  The bulk-motion 
Comptonization model predicts a cutoff in the spectrum at $E\lesssim m_{e}c^{2}$ 
(\cite{shrader98}).  In Table~1, we show that a power-law with a spectral cutoff provides 
a better fit to the high-luminosity spectrum than a power-law with no cutoff.  However, 
the measured cutoff energy of $1086^{+195}_{-145}$~keV is considerably higher than 
511~keV ($m_{e}c^{2}$).  We find that the 95\% confidence lower limit to the cutoff energy 
is 829~keV.

To explain the GRO~J1655--40 hard component, it is necessary to explain the lack of a cutoff 
in the spectrum below about 800~keV and the correlation between flux and power-law index.  
The flux-index correlation is commonly seen in AGN (\cite{grandi92};~\cite{guainazzi96}), 
and has been explained by inverse Compton scattering of soft photons on hot electrons with a 
thermal velocity distribution (\cite{haardt97}) or on electrons with a non-thermal velocity 
distribution (\cite{yaqoob92}).  However, it should be noted that the typical photon index 
observed for AGN is significantly harder than for GRO~J1655--40 and that the high energy
cutoffs observed in AGN are well below $m_{e}c^{2}$.

While the behavior of the power-law component does not depend on whether Model 1 or 
Model 2 is used, the change in the soft component does depend on which model is used.  
As shown in Table 4, if Model 1 is used, the flux of the disk-blackbody component is about 
a factor of two higher in the intermediate-luminosity state than in the high-luminosity 
state.  If Model 2 is used, the flux of the Comptonization component does not change 
significantly between states (cf. Table 5).

Although Model 1 and Model 2 provide acceptable fits to both spectra, the models are
based on somewhat different physical assumptions.  The main difference is that while Model 1 
assumes that we are seeing blackbody emission from an optically thick disk directly, Model 2 
assumes that we see the soft emission from the disk only after it is Comptonized, possibly 
in an accretion disk corona.  The Comptonization parameters we give in Table~5 assume a disk 
geometry for the Comptonization region.  In the following, we first discuss the results assuming 
Model 1 correctly describes the physical situation, then we will discuss the results assuming 
the physical situation is correctly described by Model 2.

\subsection{Model 1 Implications}

From analysis of the August 1995 GRO~J1655--40 spectrum, which is similar to our 
intermediate-luminosity state spectrum, Zhang, Cui, \& Chen~(1997b) suggest that 
the black hole in GRO~J1655--40 is rotating at about 93\% of its maximal rate in the 
prograde direction (i.e. $a_{*} = 0.93$).  We define $a_{*} = a/r_{g}$ where
$a = J/Mc$, $r_{g} = GM/c^{2}$, and M and J are the mass and the angular momentum
of the black hole, respectively.  In their calculation, Zhang et al.~(1997b) assume that 
the disk extends to the marginally stable orbit, $r_{ms}$.  When Model 1 is used to fit 
the GRO~J1655--40 spectra, the disk-blackbody parameters imply that the inner edge of the 
disk is closer to the compact object in the high-luminosity state than in the 
intermediate-luminosity state.  This may indicate that the inner edge of the disk in the 
intermediate-luminosity state does not reach $r_{ms}$.  Also, since $r_{ms}$ decreases with 
increasing black hole rotation rate, our spectral results for the high-luminosity state 
imply that $a_{*} > 0.93$.  The inner disk radius ($R_{in}$) can be derived from the 
disk-blackbody normalizations ($N_{DBB}$) using equation 1.  For the high-luminosity 
state spectrum, if $i$ = 69.5\arcdeg, then $R_{in}$ = $10.9\pm 2.6$~km.  The value of 
$r_{ms}$ for a maximally rotating $7M_{\sun}$ black hole is 10.4~km so that the value 
of $R_{in}$ is consistent with a maximally rotating black hole (\cite{bardeen72}).  

We have considered the possibility that $R_{in}$ is being underestimated because the
inclination angle of the disk close to the black hole is larger than 69.5\arcdeg.
Although the binary inclination angle has been measured by Orosz \& Bailyn (1997) to very 
high accuracy, $i = 69.50\pm 0.08$ degrees, it is possible that the inclination angle of the
accretion disk close to the black hole is different from the binary inclination.  
If we assume that the inclination of the radio jets, 85\arcdeg~(\cite{hjellming95}), gives 
the disk inclination close to the black hole, then the implied value of $R_{in}$ increases 
to $21.9\pm 5.2$~km.  Assuming that $r_{ms} = R_{in}$ gives $a_{*} = 0.93^{+0.05}_{-0.08}$.

Timing analysis has been performed for both of the RXTE observations of GRO~J1655--40 
discussed in this paper, and a complete description of this analysis will be 
published separately (\cite{remillard98}).  A QPO at 300~Hz was observed when the source
was in the high-luminosity state but not when it was in the intermediate-luminosity state.
The presence of the 300~Hz QPO in the high-luminosity state may signify an important 
difference between the states.  While, at present, there is no consensus regarding the 
physical mechanism generating the QPO, most of the suggested mechanisms require an
accretion disk terminated near the marginally stable orbit.  If 300~Hz corresponds 
to an orbital period around a $7M_{\sun}$ black hole (\cite{remillard97}), then the 
implied radius of the orbit is 64~km.  This radius (64~km) is considerably more than 
the value of $R_{in}$ implied by our spectral results, which probably indicates that 
300~Hz is not equal to the frequency of a particle orbit at the inner edge of the disk.
Other mechanisms have been suggested which may be able to explain the presence of the 
300~Hz QPO in GRO~J1655-40.  The QPO in GRO~J1655-40 may be caused by g-mode 
oscillations of the accretion disk (\cite{nowak92}) or by frame-dragging (\cite{cui98a}).  
Frame-dragging may be able to produce QPOs by causing precession of the inner region of 
the accretion disk (\cite{sv98}).

.  For both of these mechanisms, the predicted QPO frequency depends 
on the black hole rotation rate.  Coincidentally, in both cases, a value of $a_{*} = 0.95$ 
is derived when the models are applied to GRO~J1655-40 (\cite{zhang97b};~\cite{cui98a}).  
The high black hole rotation rate is in rough agreement with our spectral results.

\subsection{Model 2 Implications}

The physical picture for Model 2 is that the soft emission from the disk is Comptonized in 
a plasma near the disk.  The high-luminosity state fit parameters imply a plasma with 
a temperature of $2.66^{+0.42}_{-0.31}$~keV and an optical depth of $7.64^{+0.75}_{-0.65}$, 
which corresponds to a Comptonization y-parameter of 1.22.  After the transition to 
the intermediate-luminosity state, the plasma temperature decreases to 
$1.260^{+0.021}_{-0.026}$~keV, and the optical depth increases to $14.11^{+1.02}_{-0.70}$, which 
corresponds to y = 1.96.  If the emission responsible for the iron line is Comptonized, then 
a broad iron line, as we observe in the GRO~J1655--40 spectrum, is expected (\cite{st80}).
From the expression, $\Delta E = (E^{2}~\tau^{2})/m_{\rm e}c^{2}$ (\cite{st80}), the width 
of the high-luminosity state iron line gives $\tau \sim 5.0$.  This is in reasonably good 
agreement with the value for $\tau$ found from the Comptonization fit parameters for the 
high-luminosity state.  The centroid energy of the iron line is expected to be redshifted due 
to the recoil of the electron since the plasma temperature of the Comptonizing region is less 
than the energy of the line.  Our measured centroid energy is consistent with a redshifted 
6.4~keV iron line but the uncertainty is too large to make a meaningful redshift estimate.  
There is some precedent for broad iron lines in black hole candidates.  The mass of the 
compact object in 4U~1543-47 is between 2.5 and $7.5~M_{\sun}$ (\cite{orosz98}), indicating 
that it is likely that this source contains a black hole.  During EXOSAT observations of 
4U~1543-47, an iron line was detected with $\sigma_{line} = 1.15$~keV and $E_{line} = 5.93$~keV
(\cite{vwk89}).  Assming that Comptonization is responsible for the broadening of the iron
line, an optical depth between 4 and 5 is implied.

\section{Conclusions}

We find that the GRO~J1655--40 energy spectrum can change significantly on relatively short
timescales.  The luminosity changes from about 18\% of $L_{edd}$ to about 10\% of $L_{edd}$,
and the photon index of the power-law component changes from $2.72^{+0.03}_{-0.08}$ to
$2.415\pm 0.011$ between the high and intermediate-luminosity states.  Comparing the high and 
intermediate-luminosity state spectra shows a positive correlation between the power-law 
index and the flux of the power-law component, which is consistent with the previously
reported trend.  Models to explain the hard spectral component are constrained by this trend 
and the fact that power-law extends to 800~keV without a cutoff.
In the high-luminosity state, we find evidence for a broad iron line and a reflection
component.  In the intermediate-luminosity state, there is evidence for a broad iron line.
The fact that a 300~Hz QPO is observed only in the high-luminosity state should
provide additional information about the system.  

We regard the models we use for the soft-component as approximations to the actual 
physical system.  Global models of the accretion flow, which account for all of the 
accretion physics, are necessary.  Specifically, for GRO~J1655--40, it seems that the 
large changes in the power-law flux observed between the high-luminosity and 
intermediate-luminosity states must be related to the simultaneous changes in the soft 
component.  Construction of a physical model which generates all the observed emission 
components and reproduces the correlation between the various spectral components will be 
necessary for reliable extraction of the physical parameters of accreting compact objects.

J.A.T. would like to thank D.E. Gruber and the HEXTE group at UCSD for assistance with
the HEXTE deadtime correction and K. Leighly for useful comments and discussions.  We
also thank R.M. Hjellming and R. Hunstead for providing the results of GRO~J1655--40 
radio observations.


\clearpage

\begin{figure}[ht] \figurenum{1} \epsscale{1.0} \plotone{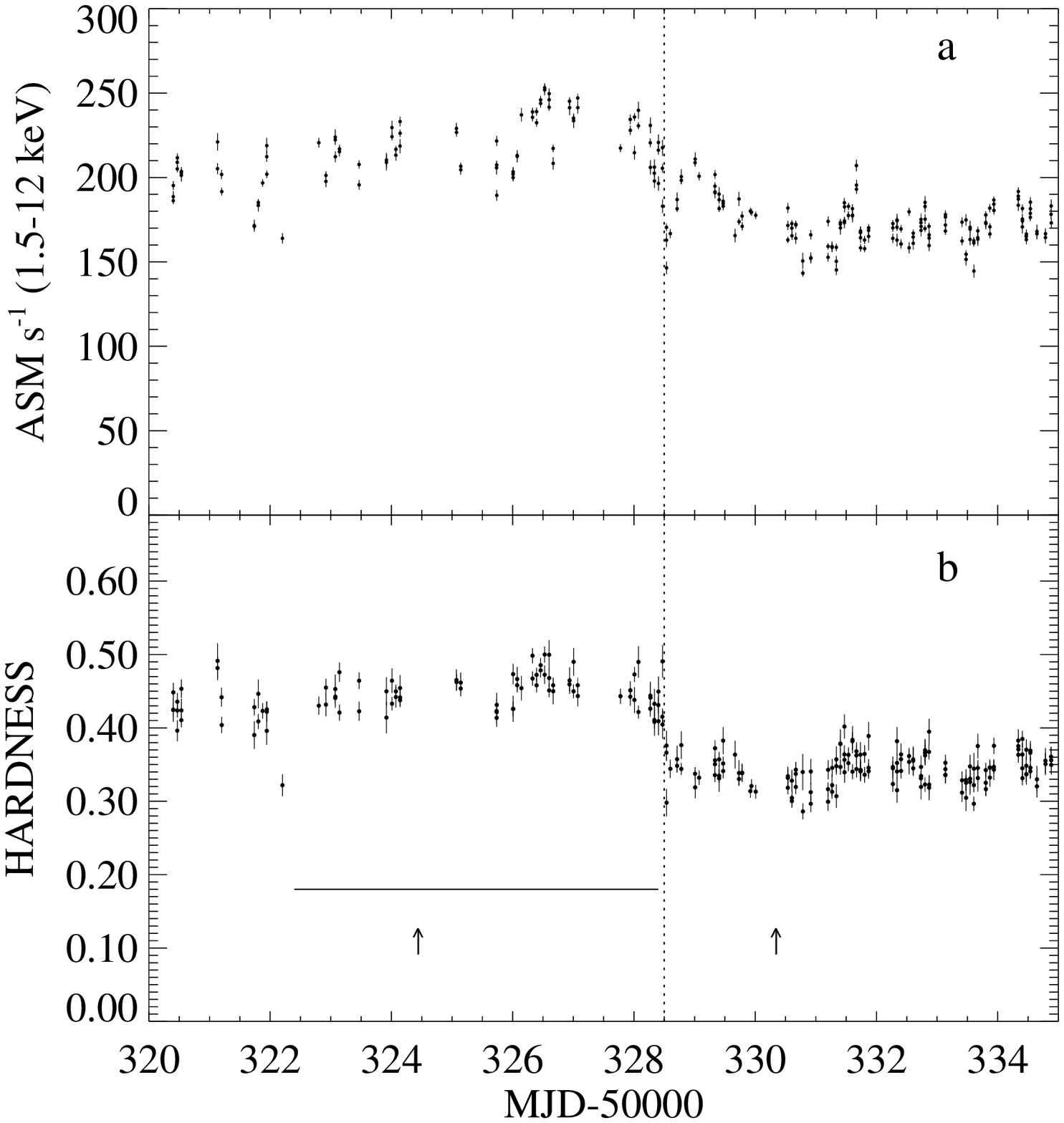}
\vspace{2.0cm}
\caption{(a) The ASM 1.5-12~keV light curve for GRO~J1655--40 and (b) the
(5-12~keV)/(1.5-5~keV) hardness ratio vs. time.  The solid horizontal line marks the
time of the OSSE observation, and the arrows mark the two pointed RXTE 
observations.  The vertical dotted line marks the transition time.\label{fig1}}
\end{figure}

\begin{figure}[ht] \figurenum{2} \epsscale{0.8} \plotone{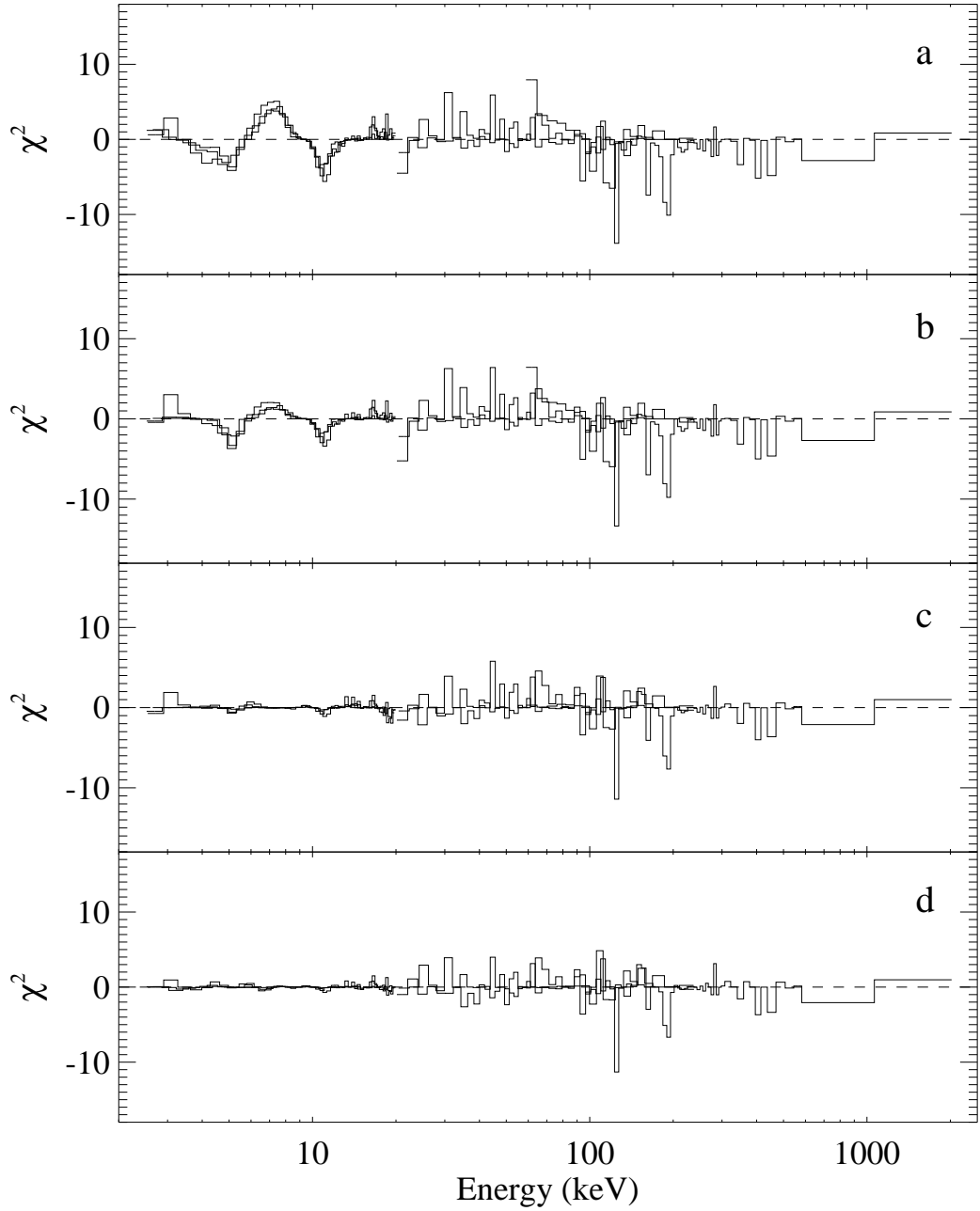}
\vspace{2.0cm}
\caption{Residuals for fits to the GRO~J1655--40 pre-transition spectrum.  
The four plots show, (a) the residuals for a disk-blackbody plus power-law fit to 
the GRO~J1655--40 spectrum, (b) the residuals for a Comptonization plus power-law fit,  
(c) the residuals for a disk-blackbody plus gaussian plus power-law plus reflection fit, 
(d) the residuals for a Comptonization plus gaussian plus power-law plus reflection fit.\label{fig2}}
\end{figure}

\clearpage
\begin{figure*} \figurenum{3} \epsscale{0.9} \plotone{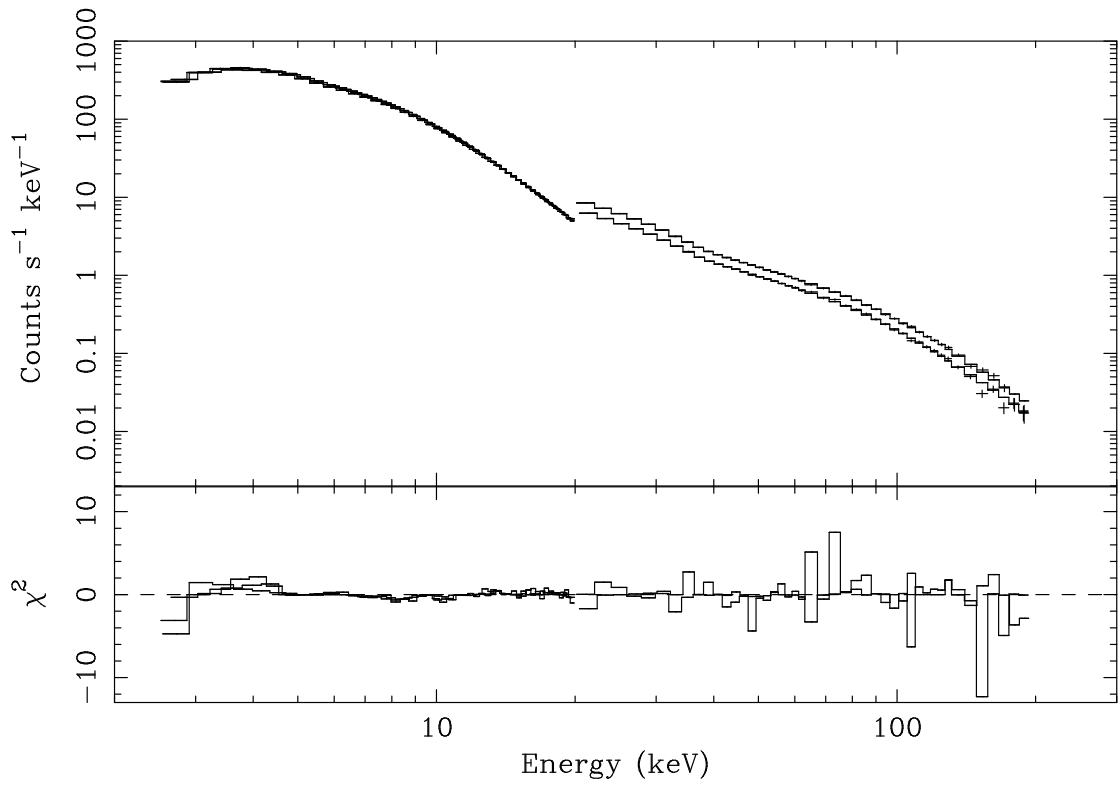}
\vspace{-5.0cm}
\caption{PCA (layer 1 only) and HEXTE Crab spectrum fit with a 
broken power-law model.\label{fig3}}
\end{figure*}

\begin{figure}[ht] \figurenum{4} \epsscale{1.0} \plotone{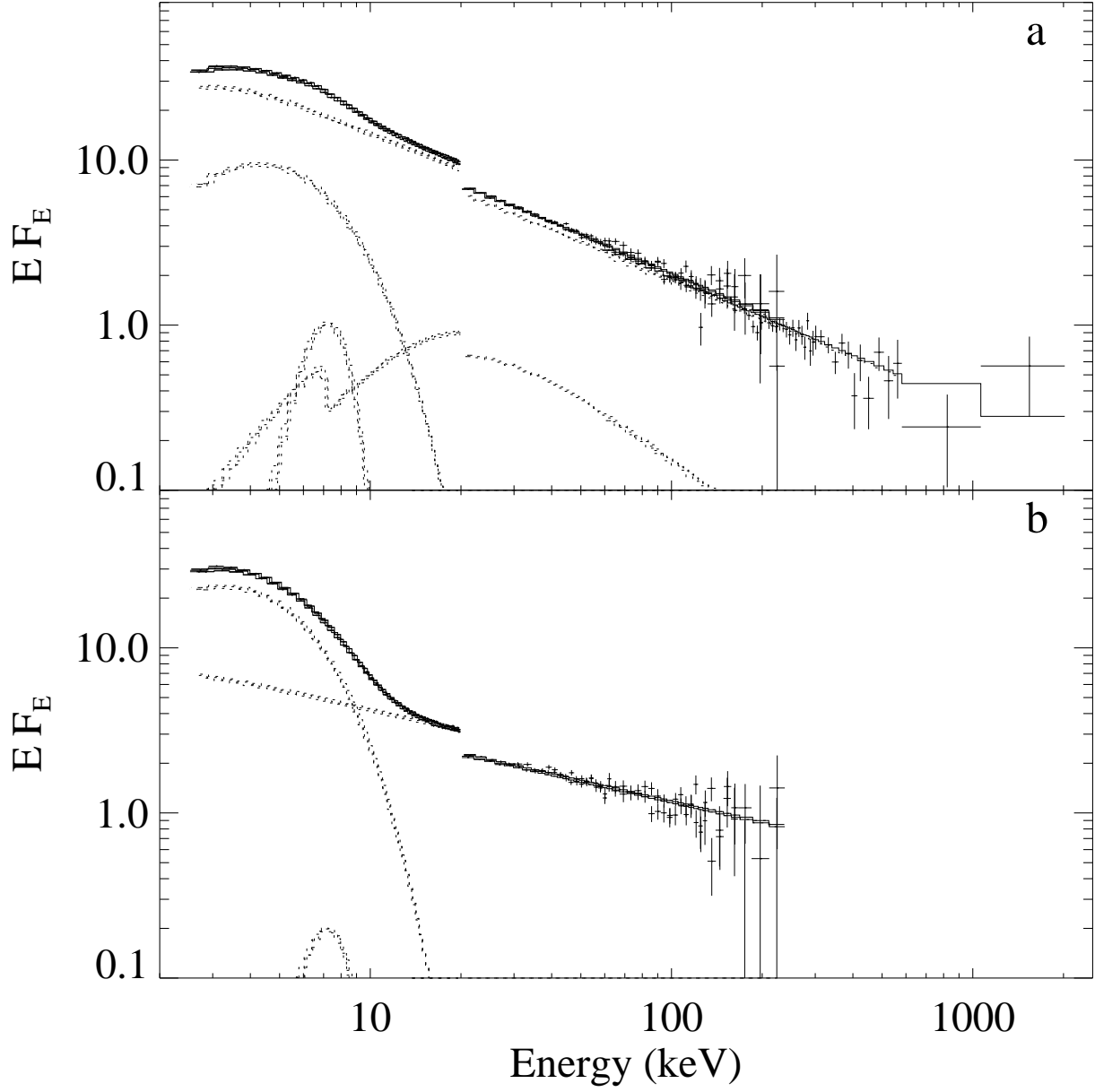}
\vspace{2.0cm}
\caption{(a) and (b) show the pre-transition and post-transition 
GRO~J1655--40 spectra, respectively.  In each plot, the solid line is the
Model 1 fit to the data, and the dashed lines show the Model 1 spectral 
components.  No normalization correction has been made.\label{fig4}}
\end{figure}

\begin{figure}[ht] \figurenum{5} \epsscale{1.0} \plotone{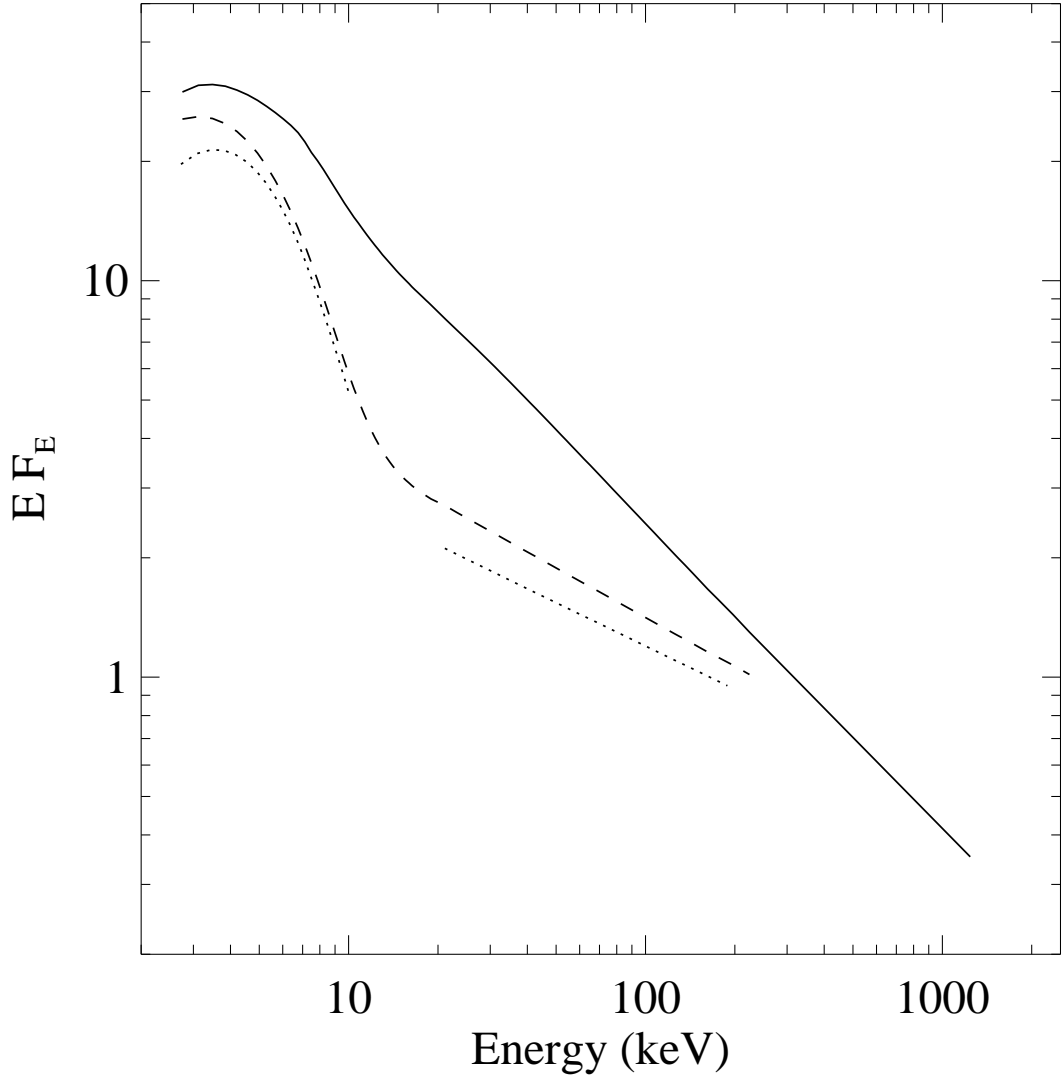}
\vspace{1.0cm}
\caption{GRO~J1655--40 broad-band spectra.  The solid line is the 
Model 1 fit to the high-luminosity spectrum, the dashed line is the Model 1 fit to 
the intermediate-luminosity spectrum, and the dotted line is the spectrum from an 
August 1995 ASCA and BATSE observation of GRO~J1655--40 based on the fit parameters 
given in Zhang et al. (1997a).\label{fig5}}
\end{figure}


\clearpage

\begin{deluxetable}{lc}
\footnotesize
\tablecaption{Fits to the HEXTE and OSSE Pre-Transition Spectrum \label{tbl-1}}
\tablewidth{0pt}
\tablehead{}
\startdata
Power-law & \nl
$\alpha$ & $2.758\pm 0.005$  \nl
$\chi^{2}/\nu$ & 197/140 \nl
\hline
Cutoff Power-law & \nl
$\alpha$ & $2.703\pm 0.010$ \nl
$E_{f}$ (keV) & $1086^{+195}_{-145}$ \nl
$\chi^{2}/\nu$ & 149/139 \nl
\hline
Power-law with Reflection & \nl
$\alpha$ & $2.739\pm 0.006$ \nl
$\Omega/2\pi$ & $0.396\pm 0.065$ \nl
$\chi^{2}/\nu$ & 147/139 \nl
\hline
Broken Power-law & \nl
$\alpha_{1}$ & $2.727^{+0.007}_{-0.010}$ \nl
$E_{b}$ (keV) & $63^{+9}_{-14}$ \nl
$\alpha_{2}$ & $2.818\pm 0.013$ \nl
$\chi^{2}/\nu$ & 144/138 \nl
\enddata
\end{deluxetable}

\begin{deluxetable}{lcc}
\footnotesize
\tablecaption{$\chi^{2}$ values for the disk-blackbody (DBB) fits \label{tbl-2}}
\tablewidth{0pt}
\tablehead{\colhead{Model} & \colhead{$\chi^{2}/\nu$ (Pre-Transition)} & \colhead{$\chi^{2}/\nu$ (Post-Transition)}}
\startdata
DBB+power-law & 391/274 & 187/221 \nl
DBB+power-law+reflection & 244/273 & 187/220 \nl
DBB+gaussian+power-law & 220/271 & 175/218 \nl
DBB+gaussian+power-law+reflection & 191/270 & 175/217 \nl
\enddata
\end{deluxetable}

\begin{deluxetable}{lcc}
\footnotesize
\tablecaption{$\chi^{2}$ values for the Comptonization (COMP) fits \label{tbl-3}}
\tablewidth{0pt}
\tablehead{\colhead{Model} & \colhead{$\chi^{2}/\nu$ (Pre-Transition)} & \colhead{$\chi^{2}/\nu$ (Post-Transition)}}
\startdata
COMP+power-law & 291/273 & 188/220 \nl
COMP+power-law+reflection & 229/272 & 188/219 \nl
COMP+gaussian+power-law & 218/270 & 155/217 \nl
COMP+gaussian+power-law+reflection & 171/269 & 155/216 \nl
\enddata
\end{deluxetable}

\begin{deluxetable}{lcc}
\footnotesize
\tablecaption{GRO J1655--40 spectral fits with Model 1\tablenotemark{a} \label{tbl-4}}
\tablewidth{0pt}
\tablehead{\colhead{Parameter} & \colhead{Pre-transition\tablenotemark{b}} & \colhead{Post-transition\tablenotemark{c}}}
\startdata
$N_{\rm H}$~($10^{22}\rm~H~atoms~cm^{-2}$) & $1.803^{+0.084}_{-0.089}$ & $0.372\pm 0.055$ \nl
Disk-Blackbody & & \nl
$kT_{max}$~(keV) & $1.489^{+0.023}_{-0.026}$ & $1.227\pm 0.004$ \nl
$N_{DBB}$ & $49.0\pm 2.7$ & $245.8\pm 4.2$ \nl
Flux~(photons cm$^{-2}$ s$^{-1}$)\tablenotemark{d} & 2.39 & 5.00 \nl
Power-law & & \nl
$\alpha$ & $2.747\pm 0.006$ & $2.416\pm 0.009$ \nl
Flux~(photons cm$^{-2}$ s$^{-1}$)\tablenotemark{d} & $7.98\pm 0.23$ & $1.742\pm 0.049$ \nl 
Gaussian & & \nl
$E_{line}$~(keV) & $6.81^{+0.24}_{-0.31}$ & $6.81\tablenotemark{e}$ \nl
$\sigma_{line}$~(keV) & $1.21^{+0.20}_{-0.18}$ & $1.21\tablenotemark{e}$ \nl
$N_{line}$~(photons cm$^{-2}$ s$^{-1}$) & $0.057^{+0.022}_{-0.015}$ & $0.0107\pm 0.0036$ \nl
EW~(eV) & 113 & 38 \nl
Reflection & & \nl
$\Omega/2\pi$ & $0.259^{+0.054}_{-0.053}$ & $0.0\tablenotemark{e}$ \nl
$i$\tablenotemark{e} & 69.5\arcdeg & - \nl
$\chi^{2}/\nu$ & 191/270 & 178/220 \nl
\tablenotetext{a}{The errors are 68\% confidence for one interesting parameter ($\Delta \chi^{2} = 1.0$).}
\tablenotetext{b}{Fit includes data from the PCA, HEXTE, and OSSE.}
\tablenotetext{c}{Fit includes data from the PCA and HEXTE.}
\tablenotetext{d}{This is the unabsorbed 2.5-20 keV component flux.}
\tablenotetext{e}{Fixed.}
\enddata
\end{deluxetable}

\begin{deluxetable}{lcc}
\footnotesize
\tablecaption{GRO J1655--40 spectral fits with Model 2\tablenotemark{a} \label{tbl-5}}
\tablewidth{0pt}
\tablehead{\colhead{Parameter} & \colhead{Pre-transition\tablenotemark{b}} & \colhead{Post-transition\tablenotemark{c}}}
\startdata
$N_{\rm H}$~($10^{22}\rm~H~atoms~cm^{-2}$) & $2.99^{+0.18}_{-0.16}$ & $2.57^{+0.30}_{-0.39}$ \nl
Comptonization & & \nl
$kT$~(keV) & $2.66^{+0.42}_{-0.31}$ & $1.260^{+0.021}_{-0.026}$ \nl
$\tau$ & $7.64^{+0.75}_{-0.65}$ & $14.11^{+1.02}_{-0.70}$ \nl
Flux~(photons cm$^{-2}$ s$^{-1}$)\tablenotemark{d} & $5.7\pm 1.4$ & $6.7^{+1.2}_{-1.3}$ \nl
Power-law & & \nl
$\alpha$ & $2.691^{+0.023}_{-0.056}$ & $2.414\pm 0.010$ \nl
Flux~(photons cm$^{-2}$ s$^{-1}$)\tablenotemark{d} & $5.7^{+1.1}_{-1.9}$ & $2.050^{+0.059}_{-0.053}$ \nl
Gaussian & & \nl
$E_{line}$~(keV) & $6.21^{+0.31}_{-0.33}$ & 6.21\tablenotemark{e} \nl
$\sigma_{line}$~(keV) & $1.69^{+0.16}_{-0.15}$ & 1.69\tablenotemark{e} \nl
$N_{line}$~(photons cm$^{-2}$ s$^{-1}$) & $0.203^{+0.054}_{-0.039}$ & $0.077^{+0.013}_{-0.016}$ \nl
EW~(eV) & 319 & 192 \nl
Reflection & & \nl
$\Omega/2\pi$ & $0.85^{+0.68}_{-0.25}$ & 0.0\tablenotemark{e} \nl
$i$\tablenotemark{e} & 69.5\arcdeg & - \nl
$\chi^{2}/\nu$ & 171/269 & 159/219 \nl
\tablenotetext{a}{The errors are 68\% confidence for one interesting parameter ($\Delta \chi^{2} = 1.0$).}
\tablenotetext{b}{Fit includes data from the PCA, HEXTE, and OSSE.}
\tablenotetext{c}{Fit includes data from the PCA and HEXTE.}
\tablenotetext{d}{This is the unabsorbed 2.5-20 keV component flux.}
\tablenotetext{e}{Fixed.}
\enddata
\end{deluxetable}

\end{document}